\documentclass[prb,preprint,showpacs,preprintnumbers,amsmath,amssymb]{revtex4}
\usepackage{graphicx}

\begin{document}
%


\let\a=\alpha \let\b=\beta \let\g=\gamma \let\d=\delta
\let\e=\varepsilon
\let\z=\zeta \let\th=\theta\let\k=\kappa \let\l=\lambda
\let\iu=\upsilon
\let\m=\mu \let\n=\nu \let\p=\pi \let\ro=\rho \let\c=\chi
\let\s=\sigma \let\t=\tau \let\f=\varphi \let\o=\omega
\let\ph=\phi
\let\G=\Gamma \let\D=\Delta \let\Th=\Theta \let\L=\Lambda \let\N=\nabla
\let\P=\Pi \let\Si=\Sigma\let\F=\Phi \let\Ps=\Psi \let\O=\Omega


\def\V#1{\vec#1}
\def\ve{{\V E}}


\def\E{{\cal E}} \def\ET{{\cal E}^T} \def\LL{{\cal L}} \def\VV{{\cal V}}
\def\SS{{\cal S}} \def\DD{{\cal D}} \def\GG{{\cal G}} \def\FF{{\cal F}}
\def\TT{{\cal T}} \def\QQ{{\cal Q}} \def\MM{{\cal M}} \def\OO{{\cal O}}
\def\BB{{\cal B}} \def\A{{\cal A}}


\def\bq{{\bf q}} \def\bh{{\bf h}} \def\bk{{\bf k}} \def\bp{{\bf p}}
\def\bs{{\bf s}} \def\bj{{\bf j}} \def\bv{{\bf v}} \def\bx{{\bf x}}
\def\bm{{\bf m}}

\def\bR{{\bf R}} \def\bQ{{\bf Q}} \def\bE{{\bf E}}
\def\br{{\bf r}} \def\bS{{\bf S}}
\def\bqh{\frac{\bq}{2}} \def\brh{\frac{\br}{2}}


\let\ra=\rightarrow
\let\Ra=\Rightarrow
\let\lra=\longrightarrow
\let\ran=\rangle
\let\lan=\langle
\def\RR{\hbox{\msytw R}} \def\CC{\hbox{\msytw C}}
\def\pb{\bar\psi} \def\pt{\tilde\psi}
\let\pd=\partial \def\Pd{\V\pd}
\def\nb{\nabla}
\def\tc{T_{c}}
\def\tf{\t_{\f}}
\def\up{\uparrow}
\def\down{\downarrow}
\def\Om{\O_m}
\def\on{\o_n}
\def\rTO{\sqrt{\frac{T}{\O}}}
\def\TO{\frac{T}{\O}}
\def\Vq{V_\bq}
\def\xik{\xi_\bk}

\newcommand{\tr}{\mathrm{Tr}}

\def\be{\begin{equation}}\def\ee{\end{equation}}
\def\bea{\begin{eqnarray}}\def\eea{\end{eqnarray}}
\def\ba{\begin{array}}\def\ea{\end{array}}
\def\nn{\nonumber}
\def\lb{\label}
\def\pref#1{(\ref{#1})}

\def\tende#1{\,\vtop{\ialign{##\crcr\rightarrowfill\crcr
\noalign{\kern-1pt\nointerlineskip} \hskip3.pt${\scriptstyle
#1}$\hskip3.pt\crcr}}\,}

\def\insertplot#1#2#3#4{\begin{minipage}{#2}
\vbox {\hbox to #1 {\vbox to #2 {\vfil%
\includegraphics{#4.ps}#3}}}
\end{minipage}}

\def\ins#1#2#3{\vbox to0pt{\kern-#2 \hbox{\kern#1
#3}\vss}\nointerlineskip}

\title{Quasi-particle dephasing time in disordered d-wave
superconductors}

\author{ M. Bruno$^{1}$}
\author{A. Toschi$^1$}
\author{L. Dell'Anna$^{1,2}$}
\author{C. Castellani$^1$}

\affiliation{$^1$Dipartimento di Fisica, Universit\`a di Roma
``La Sapienza'',\\
and Istituto Nazionale per la Fisica della Materia (INFM),
SMC and Unit\`a di Roma 1,
Piazzale Aldo Moro, 2 - 00185 Roma, Italy}
\affiliation{
$^2$Max-Planck-Institut f\"ur Festk\"orperforschung, D-70569 Stuttgart, 
Germany}

\date{\today}

\begin{abstract}
We evaluate the low-temperature cutoff for quantum interference $1/\tf$ 
induced in a $d$-wave superconductor
by the diffusion enhanced quasiparticle interactions in the presence of 
disorder. 
We carry out our analysis in the framework of the non-linear
$\sigma$-model which allows a direct calculation of  $1/\tf$,
as the mass of the transverse modes of the theory. Only the triplet
amplitude in the particle-hole channel and the Cooper amplitude
with $is$ pairing symmetry contribute to $1/\tf$. We discuss
the possible relevance of our results to the present disagreement
between thermal transport data in cuprates and the localization theory for
$d$-wave quasiparticles.
\end{abstract}

\pacs{74.25.Fy, 72.15.Rn}
\maketitle

\section{Introduction}

During the last ten-fifteen years overwhelming experimental evidences
for a $d-$wave pairing symmetry in high-$T_c$ superconducting
cuprates have been collected \cite{sper1}.  Such an unconventional
symmetry plays a crucial role in determining the physics of
the superconducting phase. In particular the presence of four nodes in
the $d-$wave superconducting gap $\D_\bk = \D (\cos k_x -\cos
k_y)/2$ and, consequently, the existence of gapless quasiparticles
(QP) excitations down to zero energy strongly affects the low
temperature transport of these systems. The energy spectrum
of the QP in the proximity of the $d-$wave gap four nodes has a Dirac
cone shape

\be
 E_{\bk}=\sqrt{\xik^2+\D_\bk^2} \simeq \sqrt{v_F^2k_1^2+v_\D^2k_2^2}
\lb{eqdirac}
\ee
where $\xik$ is the energy dispersion of the non interacting electrons, 
while $v_F$ and $v_\D$ represent respectively the Fermi velocity and
the slope of the superconducting gap $\D_\bk$ at the nodes.  
A weak amount of disorder originates\cite{lee} a 
finite quasiparticle elastic lifetime 
$\t_0$ and a finite density of 
states at the Fermi level $N_0= 1/(\pi^2 v_F v_\D \t_0) \,
\mbox{ln} (p_0 \t_0)$, $p_0$
 being a cutoff of order $\D$ . In the low temperature and frequency 
regime the conductivities 
of the system (both the electrical $\sigma$ and the thermal $\k$) do not depend
on the amount of disorder but only on the bare QP spectrum parameters
$v_F$ and $v_\D$. This leads to the so-called {\sl universality} of the 
low energy values of the conductivities in $d-$wave 
superconductors.
Indeed, by allowing the impurity-scattering between all the four nodes
of the $d-$wave spectrum and neglecting the vertex corrections, the
following ``universal'' expressions  are found\cite{lee}

\be
\lim_{\O,T \ra 0} \sigma(\O,T) = \frac{e^2}{\pi^2}\frac{v_F}{v_\D},\quad
 \lim_{\O,T \ra 0}
\frac{\k(\O,T)}{T}=\frac{k_B^2}{3}\frac{v_F^2+v_\D^2}{v_F v_\D}.
\label{equniv}
\ee
The inclusion of vertex corrections modifies $\sigma$ by a factor depending
on disorder and 
Fermi-Liquid parameters, while it leaves  $\k$ totally 
unchanged\cite{lee}.

However, in cuprate superconductors this generic result could be 
strongly affected by localization corrections because these  materials
are quasi-bidimensional and quantum interference effects are usually 
relevant in low-dimensional systems\cite{ramak}. 

>From a theoretical point of view the evaluation of the
localization corrections to the conductivities in a $d-$wave 
superconductor is a really complex topic. Indeed, when considering
a $d-$wave BCS Hamiltonian in the presence of disorder and neglecting 
interactions among quasiparticles, the theory for
the quantum interference predicts a multitude of different
regimes and crossovers depending on the specific symmetries of the
underlying Fermi surface and of the 
disorder\cite{senthil1,fukui,mesoscopic,ners,luca,luca2,yash}.
Anyway, in the generic case of non-magnetic disorder
connecting all nodal points the theoretical calculations indicate
the occurrence  of an Anderson localization
of the $d-$wave QP excitations\cite{senthil1}. 
The theory also predicts precursor effects, the so-called
weak localization corrections, which are logarithmic in
temperature.

The analysis of experimental findings for the thermal
transport in the cuprates does not appear to fit the above theoretical 
scenario.
Indeed when comparing these  predictions
with the experimental measurements a convincing
agreement is found neither with the ``universal behavior'' nor with
the complete localization.
It is worth stressing that thermal conductivity is conveniently used in
order to test the theoretical results,
because, differently from charge conductivity, it is
not affected by the vertex corrections allowing for a direct comparison
between the expression of $\kappa(T)$ in Eq. \pref{equniv} and the
experimental data.

Thermal conductivity measurements have been performed 
on various cuprates, down to temperatures of the order of $100\, mK$.
In particular the residual thermal
electronic conductivity $\k_{res}/T$ ($= \lim_{T\rightarrow
0}\k/T$), extrapolated from measurements carried out on
optimally doped $Y Ba_2 Cu_3 O_{6+x}$ ($YBCO(123)$)\cite{taillefer,chiao1} and 
$Bi_2 Sr_2 Ca Cu_2 O_8$ 
($BSCCO$)\cite{chiao1} seems to be in agreement with the
universal result (Eq. \pref{equniv} with $v_{F}$ and $v_{\D}$ extracted from
ARPES  measurements \cite{mesot,timusk}). The agreement is particularly
good for $BSCCO$, leaving no room for sizeable weak localization
corrections. The situation is partially different for $YBCO(123)$, where the 
inclusion of sizeable weak localization corrections would provide not 
unplausible larger ratios $ v_F/v_\D$ than the estimate of Ref.
\onlinecite{chiao1}.
However (or moreover) thermal magnetoconductance measurements\cite{magnth} 
do not give any
indications of the presence of detectable weak localization corrections
which should appear as a positive $H^2$ crossing to a $lnH$ contribution
at very small fields. This result is quite generic for all materials at
different dopings. On the other hand,
measurements performed on $Y Ba_2 Cu_4 O_8$ ($YBCO(124)$) underdoped  and 
$Pr_{2-x}Ce_{x}CuO_{4}$ ($PCCO$) optimally doped provide a residual
conductivity $\k_{res}/T$ much smaller than the universal value. Such
a result, compatible with $\k_{res}/T = 0$ could be 
ascribed to the localization of the quasi-particles responsible of 
the low temperature energy transport. Recently measurements of $\k_{res}/T$ 
at various doping became available in $La_{2-x}Sr_xCuO_4$ (LSCO, Ref. 
\onlinecite{takeya}) and in
$YBCO(123)$ \cite{sutherland} and made the theoretical scenario
even more involved. In particular a significant doping
dependence of $\k_{res}/T$ has been observed in these cuprates.
Such a dependence is hardly compatible with the universal behavior, 
particularly in the LSCO compounds, since it would imply a strong variation 
on doping 
of $v_{F}$ and $v_{\D}$ in partial contrast with ARPES 
measurements\cite{timusk}
and the temperature dependence of the penetration depth. Moreover 
$\k_{res}/T$ turns out  to be finite 
and no localization is found for all dopings in the superconducting phase.

Concerning the absence of weak localization contributions  a main issue is the
estimate of the temperature below which quantum interference
effects start to be visible before localization occurs.
 In the metallic 
phase this issue involves the determination of the dephasing time $\tf$, i.e.
the time scale above which  coherence is destroyed by inelastic processes.
The energy scale $1/\tf$ represents a cutoff for quantum interference and 
plays a crucial role in determining quantitatively
the magnitude of the localization effects at low temperature\cite{ramak}. 

In principle, the situation becomes more involved when considering a  $d-$wave superconductor.
Due to the extra symmetry of the Cooper pairing, quantum interference 
determines a relevant correction to the density of states which is logarithmic
in energy since it derive from integration on diffusive modes with energy as
a cutoff\cite{senthil1,khve}. This fact reflects in the conductivity with the onset of an other source of
logarithmic corrections coming from the thermal
average of the density of states contribution. These corrections 
no longer depend logarithmically on the dephasing time ($\propto-ln(\tf/\t_0)$), 
but directly on T similarly
to the interaction contributions\cite{ramak} ($\propto-ln(1/T\t_0)$).
As a consequence one can naturally individuate two contributions to the
conductivity corrections, coming out respectively from the modes damped by 
$\tf^{-1}$ (as in the normal metal) and from those damped by the energy     
$$
 \frac{\d\s_{s}}{\s_{s}} =  \left(\frac{\d\s_{s}}{\s_{s}}\right)_{\tf^{-1}}+
\left(\frac{\d\s_{s}}{\s_{s}}\right)_{\epsilon} 
$$
that can be viewed, roughly speaking, as the renormalization of the 
diffusion coefficient and of the density of states  
($\d\s_{s} \sim  \d D  N_0 + D  \d N_0  $). Indeed half of the 
contribution to the Renormalization Group (RG) equation\cite{senthil1,luca2} 
for $\s_{s}$ comes from the first term and half from the second term.

However it can be easily shown that in the range of temperature 
from $100 mK$ to $1 K$,  to which we are interested in, 
the second kind of corrections can be neglected\cite{nota_negl}, 
because of the rather small value of $1/T\t_0$ which is usually
much less than $\tf/\t_0$.  
We can therefore estimate  the localization correction to 
the spin conductivity $\s_{s}$ of a $d-$wave superconductor,
(and consequently to the thermal conductivity which is related
to $\s_{s}$ by a generalized Wiedmann-Franz law\cite{lee}),
considering only the expression for $(\frac{\d\s_{s}}{\s_{s}})_{\tf^{-1}}$
given by 
\be
\left(\frac{\d\s_{s}}{\s_{s}}\right)_{\tf^{-1}}
=\frac{\d \kappa_{res}}{\kappa_{res}}= -
 \frac{t}{2}\,\ln\frac{\tf}{\t_{0}}
\label{eqn:correction}
\ee
where $t=1/2\pi^{2}\s_{s}$ and $\s_s=(v_F^2+v_\D^2)/(\pi^2 v_F v_\D)$.

Note that the right term of the Eq. \ref{eqn:correction} differs for a 
factor $1/2$ from the analogous RG expression\cite{senthil1} since, 
according to the above discussion, only in half of RG contribution the 
relevant time region for the interference ranges from $\t_0$ to $ \tf$,
while in the other half it ranges from  $\t_0$ to $ 1/\epsilon\sim T$.

In normal metals the most effective mechanism of dephasing
is provided by the interactions among electrons \cite{altshuler}. 
Indeed interactions in presence of disorder 
have a twofold effect: on one side they generate
corrections which can compete with
those due to pure quantum interference, on the other they
provide an intrinsic dephasing
time $\t_{\f}$ which limits the quantum
interference processes for $T\neq 0$.

A systematic evaluation of interaction
corrections to the conductivities  in the d-wave superconducting phase
was carried out in Refs. \onlinecite{luca,luca2} (see also \onlinecite{khve}).
Since these corrections are proportional to the disorder induced
density of states $N_0$,
they should be negligible for clean enough systems and for temperatures 
of the order of $100$ mK, at which the thermal
transport measurements are presently available\cite{luca}.

The aim of this work is to extend the analysis of Ref. \onlinecite{luca}
and to derive an explicit expression for the
dephasing time $\tf$ induced in the superconducting phase  by the
interactions among the $d-$wave quasiparticles. 
In particular we will determine $1/\tf$ as the mass of the diffuson and the 
Cooperon 
(i.e., the two particle propagators in the particle-hole and 
particle-particle channels) 
in analogy with Refs. \onlinecite{FA,fukuiama,castellani1}  
where $1/\tf$ was evaluated in the metallic phase. 
Here we will perform a
one-loop analysis of the interaction terms, within the non-linear
$\s$-model which is  described in detail elsewhere \cite{luca,luca2}. 
We find that the most relevant contribution to $1/\tf$
comes from the triplet channel. Indeed  the  singlet channel contribution 
is ruled out by the lack of particle conservation in the 
superconducting state\cite{khve,luca}. 
On the other hand  the residual interaction 
in the Cooper channel, which would lead to  a $d+is$ instability 
when it is attractive, provides a subleading contribution to  $1/\tf$   
when it is repulsive, as we shall assume here.
As a final outcome the following  estimate for ratio $\tf/\t_0$, 
entering in Eq. \ref{eqn:correction} will be derived
\be
\frac{\tf}{\t_0} \propto \frac{\e_{F}}{T}(\D\,\t_0).
\label{tftzero}
\ee
where $\e_{F}$ is the Fermi energy and the
proportionality factor is of order one.
In agreement with the results for the metallic 
phase $\tf$ is proportional to the
inverse of  temperature  $1/T$. Moreover
the linear dependence of the ratio $\tf/\t_0$ on the elastic
scattering time suggests that quantum interference corrections at 
finite temperature would be larger in cleaner systems,
differently from what one could expect intuitively\cite{notatphi}.

By inserting in Eq. \pref{tftzero} the experimental
determinations of $v_F$, $v_\Delta$, $\Delta$ and $\epsilon_F$, we will
estimate the value $\tf$ for different cuprates, in order
to predict from Eq.\pref{eqn:correction} the temperature 
below which localization
corrections should become relevant. In particular in $YBCO(123)$ and
$BSCCO$, optimally doped, we will find that the corrections to
the universal results should become relevant, of order of $20\div 30$ 
per cent, for temperatures of the
order of $0.1 \div 1$ K, in disagreement with the
experimental results.  Actually, it is not clear why localization of the
quasiparticles does not occur in $YBCO(123)$,
while an explanation can be found for the bismuth cuprate $BSCCO$,
as we will explicitely point out in Sec. IV.
Indeed a rough estimate of the transport times, from electrical conductivity
at higher temperature,  suggests the existence in this last compound of 
other mechanisms for decoherence and dephasing, more efficient than the mechanism
of interaction in a disordered environment we have considered here.

Finally, in sec. V we will present the concluding remarks on the problems
posed by our findings.

\section{Field theory for disorder in the superconducting phase}

In this section we summarize the field theory approach
and introduce the non-linear $\sigma$-model (NL$\s$M), that will be
used in the following to evaluate the
dephasing time $1/\tf$ in the $d-$wave superconducting phase. 
It is worth noting that in the superconducting phase 
the evaluation of the dephasing time using the NL$\s$M is simpler and more
direct than using the standard perturbation theory, because 
of the presence  of both the normal and the 
anomalous Green functions with specific $k$ dependences.  

For the sake of clarity, we start discussing
in this section the NL$\s$M  in the absence 
of quasiparticle interactions, which will be included  
in a second step. In order to describe the properties of the
disordered system within a path integral formulation, we
introduce the spinorial representation \cite{efetov}

 \be
    \Ps(r)=\frac{1}{\sqrt 2}
    \left(
    \ba{c}
      \bar{c}(r)\\
      i\s_{y}c(r)
     \ea
     \right), \;
    \bar{\Ps}(r)=[C\Ps]^{T}.
   \ee
$\Ps(r)$ is an Nambu spinor, while $\bar{c}(r)$ and
$c(r)$ are Grassmann variables with components $c(r)_{\s,p,a}$ and
$\bar{c}(r)_{\s,p,a}$, where $\s$ refers to the spin, and $p=\pm$ is
the index of positive ($+\O$) and negative ($-\O$) frequency components.
Later $p$ will be extended to label the positive and negative 
Matsubara frequencies $\omega_{m}=\pi T m$, where $m$ is an odd integer.
$a=1...n$ is the replica index introduced by the replica trick used
to average the action over disorder, and $C$ is the charge
conjugation matrix

\be  \label{eqn:charge}
    C=i\s_{y}\t_{1}.
\ee
Here and thereafter, the Pauli matrices $\s_{i}(i = x, y, z)$
will act on the spin components, $s_{i}(i =1, 2, 3)$
on the frequency components, and $\t_{i}(i =1, 2, 3)$ on the
components of the Nambu spinors $\bar{c}_{i}$ and $c_{i}$.

In this representation the BCS action, describing the system in the
superconducting phase in the absence of disorder, reads

\be
 S_{0} =\sum_{\bk}\bar{\Ps}_{\bk}\left(\xik+i\D_{\bk}\t_{2}s_{1}
 -i\O s_{3}\right)\Ps_{\bk}
 \label{eq:bare-action}
\ee
where the  term $-i\O s_{3}$
has been introduced in order to define retarded and advanced Green functions,
and the presence of the Pauli matrix $s_{1}$ in the second term
is due to the fact that the Cooper pairing couples states with opposite
energies.

 As in the standard theory of localization in
the normal phase \cite{efetov}, disorder is introduced via an impurity scattering 
potential $V(r)$ with local  Gaussian distribution and zero mean value 
(the overline indicates the
impurity spatial average)
$$ \overline{V(r)}=0 \;\; , \;\;
 \overline{V(r)V(r^{'})}={u}^{2}\d(r-r^{'}) $$
In the following we will write $u^2=1/(2\pi N_0\t_0)$ 
as set by the  saddle point solution (see below).

The average over disorder distribution is performed
by using the replica trick method.  A four field
term is generated which couples fields with different replica indices

\be  \label{eqn:inter}
 S_{imp}=-\frac{1}{4\pi N_0\t_0}
 \int dy \left( \bar{\Ps}(y)\Ps(y)\right)^{2}
\ee

This contribution is then decoupled with a standard Hubbard-Stratonovich
transformation. Consequently we introduce an  hermitian bosonic field
$Q$, $Q$ being a matrix in the $(\sigma, \tau, p, a)$-space.
The decoupled action has the following expression
$$ S_{imp}= \frac{1}{2\t_0}\left\{\int dx \left[\frac{\pi}{4}N_0\tr\,Q^{2}
   -i(\bar{\Ps},Q\Ps)\right]\right\} $$
where $(\bar{\Ps},Q\Ps)$ represents a scalar product.

The NL$\s$M can now be derived, after an integration over the
Grassman variables, by expanding the effective
action  around the saddle point
 $Q_{sp}=1/(2\t_0)s_3$
and taking into account  
only the low energy transverse fluctuations.
As a consequence, the fluctuating fields $Q$  satisfy the 
conditions $Q^2=1, \; \; \mbox{Tr} Q = 0$ and can be written in the form
\be
\lb{eqn:vicoli}
Q=U^{-1} Q_{sp} U,
\ee
where $U$ is a unitary matrix with suitable symmetries.
We also rescale $Q \rightarrow 1/2\t_0 Q $ in order to express the action
in term of an adimensional field.

The outcome of this procedure  is the following  NL$\s$M

\bea
\lb{eqn:actionsc}
S_{sc}[Q] & = & \frac{\pi}{16}\left\{\s_{s}\int dx\,\tr(\N Q)^{2}
\right. \\ \nn
     & -  &     \left.  8\,N_0\,\O\int dx\, \tr(s_{3} Q)\right\}
\eea
where
\be  \label{eqn:sigmaesse}
     \s_{s}=\frac{(2\tau_0)^{-2}}{\pi V}
             \sum_{\bk}\frac{\vec{\N}\xi_{\bk}\vec{\N}\xi_{\bk}+
             \vec{\N}\D_{\bk}\vec{\N}\D_{\bk}}{(E_{\bk}^{2}+
             (2\tau_0)^{-2})^{2}} \simeq
             \frac{1}{\pi^{2}}\frac{v_{F}^{2}+v_{\D}^{2}}{v_{F}v_{\D}} 
\ee
is the quasi-particle conductance in the Born approximation
which can be interpreted as the spin conductance of the system  
\cite{senthil1,luca}.
The appearance of $\sigma_s$ (instead of the charge conductivity, as
in the metallic case) in front of the term $\tr(\N Q)^{2}$ in
Eq. \pref{eqn:actionsc}  individuates the inverse spin conductivity as 
the natural expansion parameter of the theory for the superconducting
phase.  

The expressions for the diffuson and the Cooperon 
(which are the massless modes of the theory showing diffusive poles) 
can be calculated by expanding the NL$\s$M action $S_{sc}$ in terms of
independent fields. The  Gaussian terms will define the bare diffuson and 
Cooperon
propagators, while higher order terms will provide the corrections
due to quantum interference. To this end it is useful writing the 
unitary transformation $U$, introduced in Eq. \pref{eqn:vicoli}, 
in an exponential form $U=\mbox{e}^{W/2}$. 
The symmetry properties of $W$ are the
following: (i) anti-hermiticity $W=-W^\dag$, (ii) invariance for
charge conjugation $C W^{T} C^{T}= W^\dag$, (iii) non commutation
with $Q_{sp}$ (i.e. with $s_3$). Because of this last property, 
 $W$ has the energy structure $W \propto \a s_1 +\b s_2$
and $Q$  can be written as

\be \lb{eqn:QvsW} Q=Q_{sp} \; \mbox{e}^{W} \ee 

To the above properties of $W$, one has to add a further
constraint when considering the superconducting phase. Indeed, the
presence in Eq. \pref{eq:bare-action} of the Cooper pairing term, which is
proportional to $\t_2 s_1$ and determines the non conservation of
the electric charge, requires the invariance condition $[U,\t_2 s_1] = 0 $.
This leads to
\be \lb{eqn:t2s1} [W,\t_2 s_1] = 0 \ee 
thus reducing the number of massless modes in the superconductor. Indeed, 
in this case, the $W$ field will be forced to be proportional in the
energy space either to $s_1$ or to $s_2$. This symmetry reduction 
plays a crucial role in the evaluation of the dephasing time, since it 
implies the
vanishing of the singlet term of the electron-electron 
interaction in the superconducting system \cite{khve,luca}, as 
explicitly discussed below.

The final outcome for $W$ fields both in metallic and in
superconducting phase is summarized in the table of the appendix 
taken from Ref. \onlinecite{luca2}. The table states whether the various 
components
of the W-fields are real or imaginary and symmetric or antisymmetric.

By inserting Eq.\pref{eqn:QvsW} in Eq.\pref{eqn:actionsc}, 
in which $\Omega s_3\rightarrow\lambda_m \omega_m$ and the traces 
are extended to many Matsubara frequencies as required 
by analysis of interacting case, 
and by taking
into account the symmetries of the $W$'s the two-particle propagators 
read\cite{luca2}

\be \lb{eqn:gauss}
\begin{array}{c} \lan W^{ab}_{\frac{S}{T},i,nm}(\bq)
W^{cd}_{\frac{S}{T}, i,rq}(\bq^{\prime})\ran  = 
(\pm)\, \frac{(1-\l_{n}\l_{m})}{2}\,\d(\bq+\bq^{\prime}) D_{nm}(q) \\
\\
\times \left(\d^{ac}_{nr}\d^{bd}_{mq}\; [\pm] \; \d^{ad}_{nq}\d^{bc}_{mr}
+  \;(-1)^{i}\left[ \;
\d^{ac}_{n -r}\d^{bd}_{m -q}\;[\pm]
\;\d^{ad}_{n -q}\d^{bc}_{m -r}\right]\right)
\end{array} \ee where $S,T$ refers to the
singlet-triplet components (i.e. $W= W_S
\sigma_0+i\,\vec{W_T}\cdot \vec{\sigma}$), $i=0,1,2,3$ is the index of the 
$\t$ matrices representing the particle-hole component 
(i.e., $W_{S/T}= W_{S/T}^0
\t_0+i\,\vec{W}_{S/T}\cdot \vec{\tau}$), $abcd$ and $nmrq$ are
replica and energy indexes respectively, and 
$\lambda_n=\textrm{sign}(\omega_n)$.
On the right side
$(\pm)$ applies depending whether the W-components are real or imaginary,
and $[\pm]$ whether they are symmetric or antisymmetric. Finally 
$D_{nm}$ is the (Gaussian) propagator

\bea \lb{eqn:diffuson} D_{nm}(q) & = & \frac{1}{4\pi N_0}
\frac{1}{Dq^{2}+|\o_{n}-\o_{m}|} \\ \nn 
&\equiv & \frac{1}{4\pi N_0} 
D(q,\o_{n}-\o_{m}) \eea
with the diffusion
coefficient $D= \s_s /(2 N_0)$.

The coefficients in $D_{nm}$ will be corrected both by higher order quantum
interference effects and by the electron-electron interactions.
Quite generally, both these effects determine a renormalization of
the diffusion coefficient. In particular, interaction effects result
into a self-energy $\Sigma(q,\Omega)$ which gives rise both to a
renormalization of the diffusion coefficient, $D \ra \tilde{D}$, 
and to a renormalization of the frequencies by a factor Z.
More interestingly  the interactions  determine also
the appearing  in $D(q,\Omega)$ of a mass term (a non vanishing 
$\Sigma(q=0,\Omega=0)$), which
represents the inverse dephasing time $\tf^{-1}$. When all  corrections are
included the renormalized propagator will then read
\be \lb{eqn:diffcorr} D_{n,m}(q)=\frac{1}{4\pi N_0}
\frac{1}{\tilde{D}q^{2}+Z\,|\o_{n}-\o_{m}|+\frac{1}{\tf}} \ee
where $\tilde{D}$ and $Z$ indicate the renormalizations 
evaluated in Refs. \onlinecite{luca,luca2}, while $\tf^{-1}$ will be explicitly
computed in  the following section.

\section{Effects of the interaction: calculation of the dephasing time}

In the metallic phase the relevant interactions in the presence of disorder
are the zero harmonic singlet and triplet amplitudes in the particle-hole 
channel
$\Gamma_s$ and $\Gamma_t$ and the singlet $s$-wave amplitude in the Cooper 
particle-particle channel. In the $d$-wave superconducting phase the  
relevant residual Cooper interaction comes from the $is$-wave channel. 
In the following
we will not consider this Cooper term which has been already analyzed in 
Ref. \onlinecite{khve}.

Indeed in the case of an additional repulsive $is$-wave Cooper
interaction, it can be shown with straightforward but 
lengthy calculation that its contribution to $1/\tf$ 
is subleading with respect 
to the triplet particle-hole contribution (Eq. \pref{eqn:tf2} 
below). This is because a repulsive coupling  in the Cooper channel 
scales to zero and no relevant corrections survive from the particle-particle
 interaction.
The situation would be completely different in the case of an attraction in the
$is$-wave Cooper channel, leading to a $d+is$ instability, 
but the analysis of this regime, considered in Ref. \onlinecite{khve},
is beyond the scope of this paper.  

Starting from $\Gamma_s$ and $\Gamma_t$, and
 using the same formal steps for the metallic phase 
\cite{finkel}, the following additional
contribution to $S_{sc}$ is obtained\cite{luca2} 
\bea
\lb{eqn:SintQ}
S_{int}[Q]=&-T&\!\!\,\frac{\pi^{2}N_0}{8}
\sum_{n,m,\o,a} \sum_{l=0,3}\int dx  \\ \nn && \times 
\left\{\G_s \,
\,\tr(Q_{n,n+\o}^{aa}\t_{l}\,\s_{0})\tr(Q_{m+\o,m}^{aa}\t_{l}\,\s_{0})
\right.\nonumber\\
\!\! && \;\;\; \left. + \G_{t}
\,\tr(Q_{n,n+\o}^{aa}\t_{l}\,\vec{\s})
\tr(Q_{m+\o,m}^{aa}\t_{l}\,\vec{\s})\right\} \nn \eea

Notice that the $Q$ matrices are diagonal in the
replica space, since the 
interactions are present at each fixed disorder configuration.

At this point we can make explicit the vanishing of the singlet
contribution in Eq.\pref{eqn:SintQ} due to the additional constraint 
\pref{eqn:t2s1} related to the absence of particle conservation in the 
superconducting phase. Indeed, as discussed in the previous
section, the presence of the superconducting gap causes a 
reduction of diffusive degrees of freedom with respect to the normal case. 
More precisely, the
charge-conjugation invariance, which stands both in normal and in
superconducting systems, implies

\bea \lb{eqn:symQn} C^t Q^t C= Q & \Rightarrow &
Q_{Si,nm}^{ab}= (-1)^{i}\,Q_{Si,mn}^{ba} \nonumber\\
& & Q_{Ti,nm}^{ab}= (-1)^{i+1}\,Q_{Ti,mn}^{ba} \eea 
expliciting the charge-conjugation invariance for the singlet ($S$) and the
triplet ($T$) particle-hole channels ($i=0,3$), while in the
superconducting case the following constraint also holds

\bea \lb{eqn:symQs} \t_{2}s_{1}Q\t_{2}s_{1} = -Q & \Rightarrow &
Q_{Si,nm}^{ab}=(-1)^{i+1}\,Q_{Si,-n-m}^{ab} \nonumber\\
& & Q_{Ti,nm}^{ab}= (-1)^{i+1}\,Q_{Ti,-n-m}^{ab} \eea
which follows from Eq.\pref{eqn:t2s1}.  

In general, if we consider the following transformation for the $Q$'s

\be \lb{eqn:trasf} Q_{nm} \ra Q_{-m-n} \ee
which is achieved by both permuting and changing the sign of the
energy indices, we can decompose every $Q$ field in terms of its symmetric 
($sym$)
and anti-symmetric ($ant$) components under the transformation
\pref{eqn:trasf}. Such a decomposition is useful, since in the
superconducting case the singlet and the triplet components of the
$Q^{aa}$-fields have definite symmetry properties under the
transformation \pref{eqn:trasf}. In particular, from Eq. \pref{eqn:symQn}
and \pref{eqn:symQs} one finds that

\be \lb{eqn:symm} Q_{S,sym}^{aa}=0; \; \; Q_{T,ant}^{aa}=0 \ee
On the other hand, because of energy conservation,  
Eq.\pref{eqn:SintQ} can be written only in terms
of the symmetric components, by rearranging the energy indices
in the  energy sum. As a result, only the triplet survives in Eq.
\pref{eqn:SintQ}.

We are now in position to evaluate the dephasing time, i.e.
the mass  of the propagators which is generated by the electron-electron
interaction. In the metallic phase this mass term comes
from a specific one loop self-energy diagram containing a finite contribution 
from a branch cut. We get the same result here.
To this end, after expanding the $Q$-field appearing in $S_{int}[Q]$ 
in term of the $W$-fields, we consider the four-W-field  term

\be \lb{eqn:SintW} \begin{array}{l} S_{int}(W) =
\, -T \,\frac{\pi^{2}N_0}{32} \sum_{n_i,m_i} \sum_{a,b,c}
\sum_{l=0,3} \sum_{\bk,\bk^{\prime},\bq} \G_{t} \\  \\
\quad \times \; \tr\,\left(\l_{n_{1}}W^{ab}_{n_{1}m_{1}}(\bk)
W^{ba}_{m_{1}n_{2}}(-\bk+\bq)\t_{l}\vec{\s}\right) \\ \\
\times \; \tr\left(\l_{n_{3}}W^{ac}_{n_{3}m_{2}}(\bk^{\prime})
W^{ca}_{m_{2}n_{4}}(-\bk^{\prime}-\bq)\t_{l}\vec{\s}\right)
\d_{n_{1}-n_{2}, n_{4}-n_{3}} \end{array} \ee

By performing the contractions of the $W-$fields as it is shown in Fig. 
\ref{fig:dephasing} and using the Gaussian propagator defined
in Eq. \pref{eqn:gauss}, we obtain the following contribution
to the self-energy

\be \lb{eqn:static} \Si_{n,m}=3\pi T (1-\l_m\l_n)\sum_{n_1} \G_{t}\,
\,(1-\l_{n_1}\l_m) D_{n_1 m} \ee
where we relabeled  $m_1=m$, $n_2=n$.
This contribution is the counterpart in the superconducting
phase of the contribution 
coming from diagram $(h)$ in Ref.\onlinecite{castellani1}.
To be specific, in the following we take $\o_m>0$ and $\o_n<0$ 

\begin{figure}[htbp]
\begin{center}
\includegraphics[width=8cm]{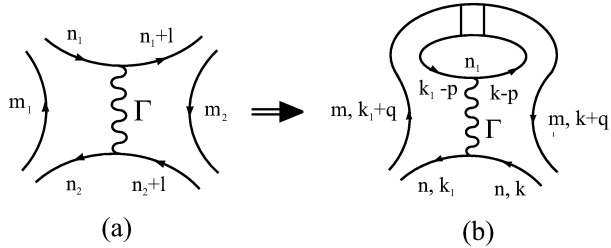}
\end{center}
\caption{(a) Four field term coming from the perturbative expansion of 
$S_{int}$ in  $W$ (Eq. \pref{eqn:SintW});  (b) Diagrammatic representation 
of the self-energy term of Eq. \pref{eqn:static}} \lb{fig:dephasing}
\end{figure}

As in the metallic phase, the dynamic structure of the interaction 
must be included in Eq.\pref{eqn:static}. Indeed the one-loop self-energy
is first order in $t$ (the disorder parameter) but its calculation
can be carried out to infinite order in $\Gamma_t$.
This is achieved simply by replacing the static
amplitude $\Gamma_t$ in Eq.\pref{eqn:static} with the dynamic amplitude
$\Gamma_t(\o)$, given by\cite{castellani1,finkel,notaonel}

\be \lb{eqn:dyn} \Gamma_t(\bq,\o)= \Gamma_{t}
\frac{D\,q^{2}+|\o|}{D\,q^{2}+(1-2\Gamma_t) \,|\o|} \ee 
where $\o=\o_{n}-\o_{n1}$. 

The self-energy $\Si$, therefore,  becomes

\be \lb{eqn:self} \Si({\bf q}=0,\Omega)= 
3\; \frac{T}{N_0}\sum_{\o_{n}<\o<\o_{n}+\O} \int
\frac{d^{d}
p}{(2\pi)^{d}}\frac{\Gamma_{t}(\bp,\o)}{D\,p^{2}+|\o+\O|} \ee
with $\Omega = \o_{m}-\o_{n}>0$.

In writing the energy limits in Eq.\pref{eqn:self} we have considered the 
partial cancellation discussed in Ref. \onlinecite{castellani1} 
leading to the upper cutoff 
$\omega<\o_{n}+\O $, so that finally the frequency sum has to be performed in 
the interval $\o_{n} <\omega <\o_{n}+\O$. Apparently this contribution
is proportional to $\Omega$ and should be taken as a contribution to Z.
However in carrying out the sum by analytically continuing on the real
frequencies $\omega \ra -i \omega$ a branch cut at $\omega=0$ appears.
Along the branch cut $\Gamma_{t}(p,\omega)$ and
$D(p,\omega+\O)$ have different analytic continuations. Because of this, the 
sum
does not vanish in the limit $\O=0$  and, in addition
to the contribution to Z, the following mass term is also obtained 

\bea \lb{eqn:tf}
\frac{1}{\tf} & = & 
\frac{2}{\pi N_0}\; \int^{\infty}_{-\infty}\, \frac{d\o}{\sinh{\b\o}} 
\\ \nn & \times & 
\int\frac{d^{d}\bp}{(2\pi)^{d}} 3 Im\{
\Gamma_{t}(\o,p)\}_{R}\,D_{R}(\o,p). \eea 
Notice that the integral in Eq. \pref{eqn:tf} is infrared divergent
and takes contribution from momentum-energy scale $0<Dq^2,\omega<T$.
As pointed out by  Fukuyama\cite{fukuiama} $1/\tf$ should enter
selfconsistlently the diffusion propagator in the  right side of
Eq. \pref{eqn:tf}. In practice, following Fukuyama we evaluate the 
integral with a low frequency cutoff of the order of $1/\tf$.
Eventually we get:

\be \lb{eqn:tf2}
\frac{1}{\tf}=2\pi\,t T \left[\frac{3}{(1-\Gamma_t)}
\Gamma_{t}^{2}\right]ln {\frac{1}{\d}}
\ee where $\d=3\,t \Gamma_t^2/(1-\Gamma_t)$ represents the
cut-off coming from the self-consistency within logarithmic accuracy.

This result indicates the existence of a direct 
proportionality between the energy scale $1/\tf$ and the temperature, with a 
coefficient of proportionality that contains the square of the scattering 
amplitude $\Gamma_t$. 
Comparing Eq. \pref{eqn:tf2}
with the results for normal metals \cite{castellani1}, 
we find that, despite the more involved Nambu formalism and the reduction
of symmetry introduced by superconductivity, the triplet contribution to
$1/\tf$ in the $d-$wave superconducting phase  has the same expression
as in the normal phase. 
Conversely, as discussed above, one finds the vanishing
of the singlet contribution, which is usually the  most relevant 
contribution in the normal phase, due to the symmetries 
of the $W$ fields in the superconducting phase. Notice also that usually
other inelastic processes (i.e. electron-phonon or electron-electron 
interactions
within the standard ``clean'' Fermi-liquid picture) lead to $1/\tf \propto
T^p$ with 
$p>1$ so that they should be subleading at low temperature with respect to 
\pref{eqn:tf2}.

\section{Size of the localization corrections and comparison with
experiments in cuprates}

In this section we use the expression \pref{eqn:tf2} to estimate the 
ratio $\tf/\t_0$ in the superconducting phase of the cuprates at low 
temperature. 
To this end we express  
the parameters appearing in Eq. \pref{eqn:tf2} in
term of experimentally accessible quantities, within few reasonable
assumptions. The parameter $\Gamma_t$ can be expressed as

\be \lb{eqn:gt} \Gamma_t= N^s_0 \, V_t \ee 
where $V_t$ is the dimensional static scattering amplitude
and for clarity we added the index $s$ to the density of 
states of the superconductor, $N^s_0$. An estimation
of $V_t$ is then obtained considering that its value should be
essentially unaffected by the superconducting transition and that,
assuming a weak- or intermediate-coupling for the metallic phase,
one has $N^m_0 \, V_t \sim 1$, i.e., $V_t \simeq 1/N^m_0$, where
$N^m_0$ is the metallic density of states. As a consequence,
being $N^s_0 \simeq 1/(\pi^2 v_F v_\D \t_0)  << N^m_0 \simeq 
k_F/(2 \pi v_F)$, 
the parameter $\Gamma_t$ will read
\be \lb{eqn:gt2} |\Gamma_t| \simeq \frac{N^s_0}{N^m_0} \simeq 
\frac{2}{\pi k_F v_\D \t_0}\simeq \frac{2}{\pi \Delta \t_0} <<1 \ee
where $\D \simeq k_Fv_\D$, being $k_F$ the Fermi momentum.

Inserting Eq. \pref{eqn:gt2} and $t=v_Fv_\D/2(v_F^2+v_\D^2)
\simeq v_\D/ (2 v_F)$  
in Eq. \pref{eqn:tf2} one gets
\be \lb{eqn:ratio}\frac{\tf}{\t_0} \simeq
\frac{\pi}{3}\frac{k_F v_F}{T}(k_Fv_\D\t_{0}) \simeq
\frac{\pi}{3}\frac{\e_{F}}{T}(\D \t_{0}) .  \ee 
where we have dropped all the logarithmic corrections and $\e_F \sim k_F v_F$.
Notice that Eq. \pref{eqn:gt2} should be considered as an upper bound for 
$\Gamma_t$ 
and consequently Eq.\pref{eqn:ratio} represents a lower bound for  $\tf/\t_0$.
By comparing with the normal phase, where the largest contribution to 
$1/\tf$ is given by the singlet interaction\cite{altshuler} and
$1/\tf \simeq \pi tT \simeq T/(k_Fv_F\t_0)$,
we find that $\tf/\t_0$ is enhanced in the superconducting phase by a 
factor $\D\t_0$, which can be quite large if disorder is weak.

At optimal doping the values for  $k_F$, $v_F$ and $v_\D$ 
can be extracted by the ARPES data, while the elastic
scattering time $\t_0$ can be deduced from optical conductivity
measurements. One gets $k_F \simeq 0.8 A^{-1}$, $v_F\simeq 1.6 eVA$, 
$v_F/v_\D \simeq 19$ (BSCCO) and $v_F/v_\D \simeq 14$ (YBCO(123))
\cite{mesot,chiao1,notaybco}.
The evaluation of  $\t_0$ is the most
problematic one. A rough estimate of $1/\t_0$ is obtained by
extrapolating to lower temperatures $1/\tau(T)$ derived from  microwave 
measurements of the charge conductivity
$\sigma(T)$ (in the temperature range of $5\div 30$ K) 
and taking  $1/\t_0=lim_{T \rightarrow 0} 1/\tau(T)$.
This leads to $1/\t_0 \simeq 0.4K$ (YBCO(123)) and 
$1/\t_0 \simeq 8K$ (BSCCO)\cite{hussey,hosseini,corson}.

From these values we obtain the following estimates for the 
dephasing time in optimally doped YBCO(123) and BSCCO 
\bea \frac{\tf}{\t_0} \; \simeq & 4\cdot 10^7 \cdot \frac{1}{T}
\;
\quad \mbox{YBCO(123)} \nonumber \\
\frac{\tf}{\t_0} \; \simeq & \; \; \; \; 10^6 \cdot
\frac{1}{T} \; \; \quad \mbox{BSCCO} \label{bscco} \eea 

We can turn now to estimate  the localization corrections to the transport 
coefficients according to Eq. \pref{eqn:correction}:

\be \lb{eqn:corr2} \frac{\d \s_s}{\s_s} =  \frac{\d \kappa_{res}}{\kappa_{res}} \simeq 
- \frac{1}{4}
\frac{v_\D}{v_F} \mbox{ln} \left[
\frac{\pi}{3}\frac{k_F v_F}{T}(k_F \, v_\D\,\t_{0})\right] \ee 

As stated before, the log factor contains  the elastic scattering 
time. As a consequence, the localization corrections 
tend -rather surprisingly- to increase with decreasing disorder.
The outcome of our calculation is that for both  materials 
we would predict that localization effects should be
visible in the temperature range we are interested in, i.e.,
hundreds of mK. In particular
for a temperature of $1$ K we get for $\d \s_s/\s_s=\d \kappa_{res}/\kappa_{res}$  
the values $-0.32$ and 
$-0.18$ in the case of YBCO(123) and BSCCO. At $100$ mK we get 
$-0.36$ and $-0.23$ respectively.
More than these specific values we want to stress that our
calculation (which possibly underestimates the effect) gives a relative 
variation of $\s_s$ and $\kappa_{res}$ of about (or even more than) 
$20 \div 30$ per cent
for temperatures lower than $1$K. 
The point is that the $t=v_\D/(2v_F)$ is small 
(but not dramatically small) while
$\tf/\t_0$ is large enough to provide a sizeable correction. For YBCO(124)
we expect similar values than YBCO(123) or even larger since these compounds
are usually quite clean.

These results are quite puzzling because, as we mentioned in the Introduction,
no relevant localization effects are found in this temperature regime but for 
the case of YBCO(124) and PCCO. 

The simplest possible explanation of such a discrepancy could be related 
to the existence in the real materials of other possible sources of inelastic 
scattering, beyond the diffusion enhanced electron-electron interaction 
(in particular scattering with some kind of quantum critical 
fluctuations\cite{qcp}), and in principle this would be responsible of
a further contribution to the dephasing effects. 

We can try to get the effective  dephasing time from the
temperature dependence of $\tau(T)$ obtained in 
Ref. \onlinecite{hosseini} and \onlinecite{corson}, as we did for $\t_0$. 
By assuming that the 
temperature dependence comes from inelastic processes and writing 
$1/\tau(T)=1/\t_0+1/\tau_{in}(T)$, we take as a rough estimate for the 
dephasing time $\tf^{exp} \simeq \tau_{in}(T)$, even though the
two times are conceptually different. This procedure gives

\bea \left(\frac{\tf}{\t_0}\right)_{exp} \; \simeq & 2\cdot 10^6
\cdot \left(\frac{1}{T}\right)^{4.2} \;
\quad \mbox{YBCO(123)} \nonumber \\
\left(\frac{\tf}{\t_0}\right)_{exp} \; \simeq & \; 10 \cdot
\frac{1}{T} \; \; \quad \mbox{BSCCO} \label{experbscco} \eea

The comparison of Eqs. \pref{experbscco} and \pref{bscco} does not reveal
any inconsistency in taking the diffusion enhanced electron-electron 
interaction 
as the predominant source of dephasing at low temperature in YBCO(123). 
In particular in the temperature range around $1K$ our previous estimate of 
$\tf$ and 
the value of $\tf^{exp}$ are of the same order of magnitude but display a 
different
temperature dependence with the ``experimental'' estimate growing as $1/T^4$ 
in the low temperature regime. As a consequence
the scattering due to the diffusion enhanced el-el interaction is
dominant below $T \simeq 0.5$ K and  our previous theoretical estimate
of Eq. \pref{bscco} has not to be corrected. 
On the contrary, in BSCCO the
``experimental'' $\tf$ displays the same $1/T$ behavior of the 
theoretical prediction, but with a much smaller coefficient, suggesting 
the presence of some other mechanism of inelastic scattering more effective 
than the
diffusion enhanced electron-electron  interaction.
In this case it becomes more appropriate to use the ``experimental'' 
estimation of
$\tf$ to evaluate the value of $\d \kappa_{res}/ \kappa_{res}$, the result
being $\d \kappa_{res}/ \kappa_{res} \simeq$ -0.03 (at 1K) and -0.05 
(at 100mK). 
These values can be taken as an indication of negligible weak localization 
corrections
in BSCCO because of strong decoherence effects.

According to the above discussion and  contrary to the experimental finding, 
significative localization effects would be expected in YBCO(123) as well as in
YBCO(124), while the presence of a different source of strong inelastic
scattering in BSCCO would explain the irrelevance of the
localization corrections even at the lowest temperatures
experimentally accessed.

\section{Conclusions}

In this paper we have dealt with the problem of the localization in 
high-$T_c$ superconductors. Their 
quasi 2-dimensional lattice structure would suggest that localization effects 
should become relevant at least in the low-temperature regime. However, as
discussed in the Introduction, the low-temperature transport measurements 
in the cuprates depend strongly on the different materials
and point out to the inadequacy of the prediction of a complete 
localization with sizeable weak localization precursor effects.  Of course the 
electronic interactions could play in principle a relevant role 
in the explanation of the real material properties. 

A main effect of the interactions, which are  a source of inelastic
scattering, is to introduce a low-energy scale, the inverse of the 
dephasing time, which limits the quantum interference processes.
Interactions also generate completely new corrections to transport,
which could in principle compete with the localization effects.
However they are expected to be negligible in the temperature-range of 
interest if disorder is weak \cite{luca}. 

We have derived the expression for the dephasing time $\tf$ in $d$-wave 
superconductors within a non-linear $\sigma$-model formulation 
of the disorder problem in the presence of interactions.
Our result differs from the result for the metallic 
phase because of the
vanishing of the singlet contribution, due to the symmetries  
of the superconducting phase. The main contribution to $1/\tf$ comes
from the interaction in the triplet channel, while 
a repulsive  residual interaction in the  Cooper
channel only  produces a subleading term. 
 
The scale $1/\tf$ allows to estimate  the size of the localization 
corrections to the heat (and spin) transport.
Indeed we have estimated the localization corrections that are theoretically 
expected  in  the various materials. We have also  considered the possibility
that other sources of inelastic scattering could provide smaller dephasing 
times
in real materials, and attempted to provide rough experimental estimates
of  $1/\tf$ to make a direct comparison with the data for thermal transport. 
The outcome of this comparison is that  the estimated dephasing time 
is not small enough to reconcile the theoretical and the experimental 
findings in the cuprates. 
This is particularly evident from the failure of the theoretical predictions 
for the 
localization corrections in the case of  YBCO(123).
Moreover, as pointed out in the Introduction, the scenario of the 
low-temperature transport in cuprates was made even more involved by very 
recent experiments in LSCO \cite{takeya} and in YBCO(123)\cite{sutherland},
which claim for a strong doping-dependence of the thermal conductivity.

The permanence of inconsistencies with the prediction of the 
localization theory, even when the electron-electron interactions 
and dephasing are taken into
account, provides a clue of the relevance of some other effects
which  have been neglected in our analysis. For instance our theoretical 
predictions could be changed substantially by the proximity to a $d+is$ 
instability 
or to a Quantum Critical Point (which would enhance considerably the estimate 
of  
the energy scale $1/\tf$), by proximity to a nesting condition, by small $q$ 
scattering 
and domain wall scattering, by the removal of the hypothesis of weak-coupling 
interactions, which could become questionable especially for the underdoped 
cuprates and large disorder. All the above effects will lower the temperature 
scale
below which localization become sizeable.

Finally we would like briefly to mention
the relevant problem of magnetic impurities in the localization
process. Since the scattering with magnetic impurities 
determines\cite{mesoscopic,luca} a vanishing of the localization 
corrections, a measurement of the low temperature transport coefficients
in presence of magnetic impurities would be of a great 
relevance for  understanding the role of the Anderson 
localization in the low-temperature transport of the cuprates, especially in 
the case of YBCO(124), 
where localization could be at work in providing 
a vanishing thermal transport. In YBCO(124) it would be also useful a 
systematic and careful analysis of magnetoconductance and thermal transport at 
small fields to compare with the findings in the cuprates which do not show 
localization. For these latter systems it would instead be important
to assess the absence of magnetic impurities and local moments by magnetic 
measurements.

\acknowledgments{
We wish to thank M.Fabrizio for valuable discussions. C.C. also
thanks P.A.Lee for useful comments on the relevance of spin flip 
scattering.}

\appendix*

\section{}

We report here explicitly the symmetries of the  
transverse massless modes W:

\begin{eqnarray*}
&&W_{S0,nm}^{ab} = \phantom{+}W_{S0,nm}^{ab*} = - W_{S0,mn}^{ba}
= \phantom{+}W_{S0,-n-m}^{ab} \\
&&W_{S1,nm}^{ab} = - W_{S1,nm}^{ab*}  = - W_{S1,mn}^{ba}
= -W_{S1,-n-m}^{ab} \\
&&W_{S2,nm}^{ab} = - W_{S2,nm}^{ab*}  = - W_{S2,mn}^{ba}
= \phantom{+}W_{S2,-n-m}^{ab} \\
&&W_{S3,nm}^{ab} = \phantom{+}W_{S3,nm}^{ab*}
=  \phantom{+}W_{S3,mn}^{ba} = -W_{S3,-n-m}^{ab} \\
&&\vec{W}_{T0,nm}^{ab} = \phantom{+}\vec{W}_{T0,nm}^{ab*}
= \phantom{+}\vec{W}_{T0,mn}^{ba} = \phantom{+}\vec{W}_{T0,-n-m}^{ab}
\\
&&\vec{W}_{T1,nm}^{ab} = - \vec{W}_{T1,nm}^{ab*}
= \phantom{+}\vec{W}_{T1,mn}^{ba} = -\vec{W}_{T1,-n-m}^{ab}
\\
&&\vec{W}_{T2,nm}^{ab} = - \vec{W}_{T2,nm}^{ab*}
= \phantom{+}\vec{W}_{T2,mn}^{ba} = \phantom{+}\vec{W}_{T2,-n-m}^{ab}
\\
&&\vec{W}_{T3,nm}^{ab} = \phantom{+}\vec{W}_{T3,nm}^{ab*}
= - \vec{W}_{T3,mn}^{ba} = -\vec{W}_{T3,-n-m}^{ab}
\end{eqnarray*}

where $n$ and $m$ are odd integers which label the Matsubara frequencies
$\omega_n=T\pi n$ and $\omega_m=T\pi m$, which have opposite sign 
($\omega_n\omega_m<0$).
Notice that because of Eq. \pref{eqn:t2s1} in the superconducting phase
$W_{nm}$ and $W_{-n-m}$ are not independent.


\begin{thebibliography}{99}
\bibitem{nota_negl} The expression of this kind of 
 corrections would read
$1/2\, t\; \mbox{ln} (\t_0 T)$, so considering that for the materials 
examined here
$t$ is always  between $1/20 \div 1/40$ and   $1/\t_0$ lower than $8$ K, 
the size  of such conductivity corrections is at most of the order of 
few percent. Notice that the factor $1/2t$ is the same factor appearing
in the energy dependent correction to the density of states
$\delta N_0/N_0 =1/2t \,ln(\t_0\epsilon)$.
\bibitem{notatphi} It has to be remarked that our results have been 
obtained in the low temperature limit and that the validity of
our theoretical description is limited to  $T\le 1/\t_0$.
\bibitem{notaonel} This result can be obtained explicitly by considering
the one loop contractions of 2 three-$W$ vertices and by inserting   
two-$W$ interaction vertices.
\bibitem{notaybco} Actually, this last estimate is derived from
the thermal conductivity assuming the universal value \pref{equniv}.
\end{thebibliography}
\end{document}